\documentclass{article}

\usepackage{arxiv}

\usepackage[utf8]{inputenc} 
\usepackage[T1]{fontenc}    
\usepackage{hyperref}       
\usepackage{url}            
\usepackage{booktabs}       
\usepackage{amsfonts}       
\usepackage{nicefrac}       
\usepackage{microtype}      
\usepackage{lipsum}		
\usepackage{graphicx}
\usepackage{natbib}
\usepackage{doi}
\usepackage{amsmath}   
\usepackage{graphicx}  
\usepackage{adjustbox} 
\usepackage{multirow}

\title{Integrating Newton's Laws with deep learning for enhanced physics-informed compound flood modeling}

\author{ \href{https://orcid.org/0000-0003-0177-9733}{\includegraphics[scale=0.06]{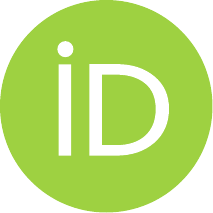}\hspace{1mm}Soheil Radfar}\\
	Center for Complex Hydrosystems Research\\
	Department of Civil, Construction \\
    and Environmental Engineering\\
    The University of Alabama\\
	Tuscaloosa, AL, USA \\
	\texttt{sradfar@ua.edu} \\
	\And
	\href{https://orcid.org/0000-0002-6522-0712}{\includegraphics[scale=0.06]{orcid.pdf}\hspace{1mm}Faezeh Maghsoodifar} \\
	Center for Complex Hydrosystems Research\\
	Department of Civil, Construction \\
    and Environmental Engineering\\
    The University of Alabama\\
	Tuscaloosa, AL, USA \\
	\texttt{fmaghsoodifar@crimson.ua.edu} \\
    \And
	\href{https://orcid.org/0000-0003-3170-8653}{\includegraphics[scale=0.06]{orcid.pdf}\hspace{1mm}Hamed Moftakhari} \\
	Center for Complex Hydrosystems Research\\
	Department of Civil, Construction \\
    and Environmental Engineering\\
    The University of Alabama\\
	Tuscaloosa, AL, USA \\
	\texttt{hmoftakhari@eng.ua.edu} \\
    \And
	\href{https://orcid.org/0000-0003-3170-8653}{\includegraphics[scale=0.06]{orcid.pdf}\hspace{1mm}Hamid Moradkhani} \\
	Center for Complex Hydrosystems Research\\
	Department of Civil, Construction \\
    and Environmental Engineering\\
    The University of Alabama\\
	Tuscaloosa, AL, USA \\
	\texttt{hmoradkhani@ua.edu} \\
}

\renewcommand{\shorttitle}{Radfar \textit{et al.}, 2025}

\hypersetup{
pdftitle={A template for the arxiv style},
pdfsubject={q-bio.NC, q-bio.QM},
pdfauthor={David S.~Hippocampus, Elias D.~Striatum},
pdfkeywords={First keyword, Second keyword, More},
}

\begin{document}
\maketitle

\begin{abstract}
	Coastal communities increasingly face compound floods, where multiple drivers like storm surge, high tide, heavy rainfall, and river discharge occur together or in sequence to produce impacts far greater than any single driver alone. Traditional hydrodynamic models can provide accurate physics-based simulations but require substantial computational resources for real-time applications or risk assessments, while machine learning alternatives often sacrifice physical consistency for speed, producing unrealistic predictions during extreme events. This study addresses these challenges by developing ALPINE (All-in-one Physics Informed Neural Emulator), a physics-informed neural network (PINN) framework to enforce complete shallow water dynamics in compound flood modeling. Unlike previous approaches that implement partial constraints, our framework simultaneously enforces mass conservation and both momentum equations, ensuring full adherence to Newton's laws throughout the prediction process. The model integrates a convolutional encoder-decoder architecture with ConvLSTM temporal processing, trained using a composite loss function that balances data fidelity with physics-based residuals. Using six historical storm events (four for training, one for validation, and one held-out for unseen testing), we observe substantial improvements over baseline neural networks. ALPINE reduces domain-averaged prediction errors and improves model skill metrics for water surface elevation and velocity components. Physics-informed constraints prove most valuable during peak storm intensity, when multiple flood drivers interact and reliable predictions matter most. This approach yields a physically consistent emulator capable of supporting compound-flood forecasting and large-scale risk analyses while preserving physical realism essential for coastal emergency management.
\end{abstract}

\keywords{Physics-informed \and Coastal flooding \and Compound flooding \and PINN \and Shallow water equations}

\section{Introduction}
Flooding is one of the most destructive and costly natural hazards worldwide, leading to widespread environmental, economic, and social repercussions. Approximately 1.8 billion people globally are exposed to 1-in-100-year flood events, with flood-related damages exceeding \$1 trillion USD since 1980 \citep{green2025comprehensive}. In the United States alone, annual flood-related economic losses range from \$179.8 to \$496.0 billion, highlighting its significant impact on infrastructure, homes, and livelihoods \citep{USCongress2024}. Coastal regions are particularly vulnerable due to dense populations and concentrated economic activities in low-lying areas, with nearly 40\% of the global population living within 100 km of coastlines \citep{Radfar2024a}. Rapid urbanization, sea-level rise, and climate change exacerbate the risk of coastal flooding, driving the need for improved mitigation and risk assessment strategies \citep{Ali2025, green2025comprehensive}.

Among coastal hazards, compound flooding (CF), which arises from the concurrent or successive occurrence of multiple flood drivers such as heavy rainfall, river discharge, and storm surge, has gained attention for its disproportionate impacts compared to univariate flood events \citep{Moftakhari2017, Zscheischler2020}. Studies indicate that CF contributes to a significant portion of coastal flood events, particularly in regions exposed to tropical cyclones, where the interplay of surge and rainfall can amplify damages \citep{Sarhadi2024, Wahl2015}. For instance, along the U.S. Gulf and East coasts, approximately 80\% of recorded flood events are compound in nature, with property losses far greater than those caused by single-driver floods \citep{Ali2025, sohrabi2025analyzing, gori2020tropical}. These events are projected to increase in frequency and intensity as climate change intensifies flood drivers, underscoring the urgent need to develop accurate and scalable models for CF hazard mapping \citep{gori2020assessing, green2025comprehensive, grimley2024climate, Radfar2024a}.

CF modeling remains one of the key challenges in coastal and hydrodynamic research due to the intricate interplay of multiple drivers \citep{Feng2024, Munoz2024}. Traditional hydrodynamic models, while robust and physically grounded, are computationally expensive, especially when simulating large-scale CF events at high spatial and temporal resolutions \citep{Munoz2021, Sarhadi2024}. Achieving the fine resolution necessary to capture rapid flood dynamics requires massive computational resources, often restricting their applicability in real-time emergency scenarios or for large-scale risk assessments \citep{Liu2024}. Moreover, such models encounter difficulties in accounting for the non-linear and simultaneous interactions between the different flood drivers, leading to uncertainties in predicting CF impacts \citep{abbaszadeh2022perspective, Munoz2024, Munoz2021, Radfar2024a}.

To address these challenges, data-driven approaches, particularly machine learning and deep learning methods, have emerged as efficient alternatives for CF mapping. Convolutional neural networks and data fusion techniques, for instance, have been successfully applied to post-event flood mapping by leveraging satellite imagery, radar data, and digital elevation models \citep{Bentivoglio2022, Munoz2021}. These approaches significantly reduce computational costs while maintaining high accuracy, making them ideal for large-scale and real-time flood assessments. However, despite their promise, purely data-driven models often fail to incorporate the underlying physical laws governing flood dynamics, which can limit their generalizability to unseen scenarios and introduce errors when extrapolating to new regions or conditions \citep{Bentivoglio2022, Donnelly2024}.

Physics-Informed Neural Networks (PINNs) offer a promising hybrid solution by combining the strengths of traditional physics-based models and modern data-driven techniques. PINNs integrate physical constraints, such as conservation laws, directly into the neural network loss function, ensuring predictions remain consistent with the governing hydrodynamic equations \citep{Raissi2019}. Recent studies have demonstrated that PINNs outperform purely data-driven approaches, particularly when working with sparse data or complex boundary conditions \citep{Donnelly2024, Liu2024}. By embedding the physics of shallow water equations into the modeling framework, PINNs provide a more accurate and computationally efficient alternative for flood simulations while preserving physical realism. As such, PINNs are well-suited to address the complexities of CF characterization, offering scalable solutions for real-time hazard prediction and risk management.

Several recent efforts have explored the integration of PINNs with shallow water dynamics, often by enforcing only part of the governing equations. \citet{Donnelly2024} incorporated volume consistency as a regularization term but did not constrain momentum. \citet{Taghizadeh2025} proposed a graph-based PINN for flood forecasting that also targeted mass balance, yet their framework omitted full momentum dynamics. A few recent studies, such as \citet{Qi2024}, \citet{Zhou2025}, \citet{Dazzi2024}, and \citet{Feng2023} began embedding components of the shallow water system in testbed settings, but their applications were limited to idealized domains, reduced-dimensional configurations, or single-source flood mechanisms.

This study builds on these developments by applying PINNs to real-world CF scenarios and develops ALPINE (All-in-one Physics Informed Neural Emulator), which to the best of the authors' knowledge, is the first to enforce all three fundamental equations of the shallow water system (i.e., the continuity equation and both momentum equations) in the context of CF modeling. By requiring the model to satisfy mass conservation together with both horizontal and longitudinal momentum equations, we frame the learning problem around the full structure of Newton’s laws. This approach avoids the risk of solutions that honor one equation at the expense of others, providing a foundation that is not only physically consistent, but also essential when modeling interactions between riverine and coastal forcing where dynamics are tightly coupled. In this way, the work advances the use of PINNs from partial constraint surrogates toward fully governed systems that respect the underlying physics in both principle and practice.

\section{Materials and methods}
\label{sec:headings}
\subsection{Study area and scenarios}
The Mobile Bay estuarine system provides an ideal testbed for CF modeling due to its unique hydrodynamic characteristics and complex interactions between riverine discharge, storm surge, and precipitation. As the fourth largest estuary by discharge in the United States, Mobile Bay receives massive freshwater inputs from the sixth largest river system in the country, which drains approximately 114{,}000~km\textsuperscript{2} across Alabama, Mississippi, Georgia, and Tennessee through the Mobile and Tensaw Rivers and their tributaries. Despite covering 1{,}070~km\textsuperscript{2}, the bay maintains a remarkably shallow average depth of only 3~m, significantly influencing its hydrodynamics and sensitivity to storm surge. The study region is particularly susceptible to intense tropical cyclones \citep{Radfar2024b}. 

To capture these dynamics, this study employs six distinct hurricane and tropical storm events spanning over a decade to reflect diverse CF scenarios with varying intensities and characteristics. The events include Hurricane Isaac (2012), Hurricane Nate (2017), Hurricane Michael (2018), Tropical Storm Claudette (2021), Hurricane Ida (2021), and Hurricane Francine (2024), as detailed in Table~\ref{tab:stormevents}. These events represent a range of storm categories from tropical storms to Category 5 hurricanes, with total damages ranging from \$295~million to \$86.7~billion.

Isaac and Nate were primarily surge-driven storms that pushed water levels to extreme heights of 1.4~m and 2.0~m, respectively, at key monitoring stations, causing significant pier damage along the eastern shore \citep{Berg2013, Beven2018}. Yet Isaac also brought $\sim$360~mm of rainfall to Grand Bay. Claudette and Francine were mainly rainfall-dominant storms. Claudette generated relatively modest storm surge of only 1.1~m at Bayou La Batre but delivered substantial precipitation with $\sim$230~mm falling in northern and central Alabama \citep{Papin2022}. Francine showed a similar pattern, with gauges topping out at 0.3--0.9~m but Danville recording $\sim$350~mm of rainfall \citep{Bucci2025}. Ida was a more balanced CF scenario with both significant storm surge of 1.0~m on the West Fowl River and widespread heavy rainfall of 200--280~mm across Mobile and Baldwin counties \citep{Beven2022}. At the low end, distant Hurricane Michael produced only minor (\textless0.6~m) water-level anomalies and less than 150~mm of rain statewide \citep{Hagen2019}. 

The most damaging coastal floods in Mobile Bay occur when a hurricane’s center passes just west of the bay, keeping the estuary in the storm’s right-front quadrant (e.g., Isaac, Nate, Ida), while east-tracking (Michael) or faster-moving storms (Francine) yield lower surge but can still deliver substantial pluvial flooding hazards. Collectively, these six storms provide a balanced benchmark set for evaluating CF risk and modeling in Alabama’s Mobile Bay watershed.

\begin{table}[htbp]
\centering
\caption{Historical events used for model training and testing. The table shows key characteristics including hurricane season, maximum wind speed, landfall area, category at landfall, total damage, and simulation period for each event. Damage values are based on official National Hurricane Center (NHC) reports and adjusted for inflation using the Consumer Price Index (CPI).}
\label{tab:stormevents}
\begin{tabular}{lcccccc}
\toprule
\textbf{Event name} & \textbf{\shortstack{Hurricane\\season}} & \textbf{\shortstack{Maximum\\wind speed\\(kt)}} & \textbf{\shortstack{Landfall\\state}} & \textbf{\shortstack{Category\\at landfall}} & \textbf{\shortstack{Total Damage,\\Adjusted\\(M US\$)}} & \textbf{\shortstack{Simulation\\period}} \\
\midrule
Hurricane Isaac            & 2012 & 70  & Louisiana   & H1 & 3290  & 08/22 -- 09/03 \\
Hurricane Nate             & 2017 & 80  & Mississippi & H1 & 295   & 10/01 -- 10/12 \\
Hurricane Michael          & 2018 & 140 & Florida     & H5 & 31973 & 10/04 -- 10/15 \\
Tropical Storm Claudette   & 2021 & 40  & Louisiana   & TS & 445   & 06/14 -- 06/25 \\
Hurricane Ida              & 2021 & 130 & Louisiana   & H4 & 86697 & 08/22 -- 09/03 \\
Hurricane Francine         & 2024 & 90  & Louisiana   & H2 & 1330  & 09/04 -- 09/15 \\
\bottomrule
\end{tabular}

\vspace{1mm}
\begin{flushleft}
\footnotesize
\textit{* The damages are based on the official reports published by the NHC for each event. The values are adjusted according to 2025 CPI provided by the U.S. Department of Labor Bureau of Labor Statistics to account for the inflation rate (source: \href{https://www.usinflationcalculator.com}{https://www.usinflationcalculator.com}).}
\end{flushleft}
\end{table}

\subsection{Hydrodynamic model}
A range of hydrodynamic models are available for simulating coastal flooding, from complex three-dimensional models to simplified two-dimensional approaches. While detailed 3D models can provide comprehensive hydrodynamic information, they often require significant computational resources and long simulation times, making them impractical for large-scale applications or multiple scenario analyses. Two-dimensional models based on shallow water equations offer a good compromise between computational efficiency and physical accuracy for coastal flooding applications. One such model that has gained attention for its computational efficiency while maintaining physical accuracy is the Super-Fast INundation of CoastS model (SFINCS; \citealp{Leijnse2021}). SFINCS is a reduced-complexity hydrodynamic model designed for computationally efficient simulations of pluvial, fluvial, and coastal flooding (e.g., \citep{sebastian2021hindcast, grimley2025determining}. It solves the shallow water equations (SWE) using a first-order explicit scheme, with the flexibility to include or neglect advection terms depending on the scenario requirements \citep{Leijnse2018}. In this study, the SFINCS-SSWE configuration is used, which includes the advection term to better capture wave-driven processes. The SSWE (or, simplified SWE) formulation (see Section~2.3 Governing equations) is achieved by neglecting the horizontal viscosity terms in the full SWE, and is the inviscid version of the Saint-Venant equations in 1D \citep{SaintVenant1871}.

This study integrates multiple publicly available geospatial datasets to construct and validate a hydrodynamic model of the Mobile Bay estuarine system. The datasets encompass topography, bathymetry, land cover, meteorological forcing, and hydrographic characteristics. The model’s terrain representation combines two complementary elevation datasets. For terrestrial and nearshore areas, we utilize the NOAA Continuously Updated Digital Elevation Model (CUDEM) at 3-meter resolution \citep{Amante2023} (available at the NOAA’s Data Access Viewer: \url{https://coast.noaa.gov/dataviewer}). Where CUDEM data are unavailable, the General Bathymetric Chart of the Oceans (GEBCO; \citealp{Weatherall2015}) supplies global coverage at a 450-meter resolution (available at: \url{https://www.gebco.net}).

Meteorological forcing for the simulations is derived from the ERA5 reanalysis dataset \citep{Hersbach2020}. The dataset is available at the Copernicus Climate Data Store (\url{https://cds.climate.copernicus.eu}) and provides hourly values of 10-meter wind components (u10, v10), mean sea level pressure, and total precipitation at 30-km spatial resolution. These inputs ensure the model captures the dynamic interactions between atmospheric and hydrological systems during extreme events. Land cover data from the ESA World Cover v2 dataset \citep{Zanaga2022}, with a spatial resolution of 10 meters (available at: \url{https://viewer.esa-worldcover.org}), are used to classify the study area into 11 distinct land cover types, including urban areas, various vegetation types, and water bodies. These classifications are mapped to Manning’s ($n$) roughness coefficients (Table~\ref{tab:manning}) for accurate representation of surface resistance in the hydrodynamic model.

\begin{table}[htbp]
\centering
\caption{Manning’s roughness coefficients ($n$) used for the hydrodynamic model of the Mobile Bay}
\label{tab:manning}
\begin{tabular}{{lc}}
\toprule
\textbf{Land cover category} & \textbf{Manning's $n$ value} \\
\midrule
Tree cover               & 0.12  \\
Shrubland                & 0.05  \\
Grassland                & 0.034 \\
Cropland                 & 0.037 \\
Built-up                 & 0.1   \\
Bare / sparse vegetation & 0.023 \\
Snow and Ice             & 0.01  \\
Permanent water bodies   & 0.02  \\
Herbaceous wetland       & 0.035 \\
Mangroves                & 0.07  \\
Moss and lichen          & 0.025 \\
\bottomrule
\end{tabular}
\end{table}

Considering infiltration conditions guides the model toward better modeling of rainfall-runoff processes in compound flood \citep{maymandi2022compound}. For this study, the infiltration effect is considered using the curve number method. This method accounts for the surface runoff and infiltration dynamics by assigning curve numbers based on land cover, soil type, and hydrologic conditions. The infiltration rates are critical for accurately simulating pluvial flooding, particularly during intense precipitation events. In the developed SFINCS models, the global average curve number values are adopted from \citep{Jaafar2019}. Hydrographic features, including flow direction, flow accumulation, and river channel characteristics, are defined using the MERIT Hydro dataset \citep{Yamazaki2019}. This dataset provides a high-resolution representation of riverine systems and is used to delineate the inflow and outflow conditions required for riverine flooding simulations.

Moreover, coastal water level forcing at the ocean boundary is implemented using hourly water level data from Dauphin Island (NOAA station ID: 8735180), which serves as a representative proxy for coastal water level variability in the region. This methodology follows the validated approach of \citep{Munoz2022} for comparable coastal conditions. The developed hydrodynamic models apply these boundary conditions at designated point locations, with SFINCS automatically interpolating the values to the corresponding water level boundary cells \citep{Eilander2023} to ensure capture water level variations at the open boundary. River flow data for the upstream boundary conditions are obtained from the U.S. Geological Survey (USGS) gauges. The model is forced with six riverine boundary conditions from major rivers and creeks in the region. These include the Mobile River (USGS station ID: 02470629), Tensaw River (02471019), Chickasaw Creek (02471001), Fowl River (02471078), Fish River (02378500), and Magnolia River (0237830). Discharge data from these stations provide crucial inputs for simulating riverine inflows and their interactions with tidal and surge processes.

A curvilinear mesh with a horizontal resolution of 200 meters was generated for the hydrodynamic model, with all water levels referenced to Mean Sea Level (MSL). Six distinct simulation periods were implemented to capture coastal dynamics across multiple timeframes within hurricane seasons (June 1 through November 30) and varying storm intensities (Table \ref{tab:stormevents}). Each simulation spans approximately 10-12 days to ensure adequate model spin-up and capture the complete temporal evolution of each storm event. The model domain is defined by a bounding box with geographic coordinates extending from [-88.52, 29.98] to [-87.41, 31.07], encompassing the Mobile Bay basin and its surrounding areas. All input datasets, if required, are reprojected and resampled to the UTM Zone 16N coordinate system (EPSG: 32616) to ensure consistency in the hydrodynamic model configuration. The study area consists of 536 grid points in the x-direction and 608 grid points in the y-direction, resulting in a total of 325,888 computational grid cells.

\subsection{Governing equations}
SWE represents a mathematical framework that directly embodies Newton's fundamental laws of motion in fluid form. The classical formulation consists of three fundamental balance equations:

\begin{equation}
\begin{adjustbox}{max width=\textwidth} 
$\displaystyle
\begin{aligned}
\text{(continuity)}\;&
  \underbrace{\partial_t h}_{\text{local inertia}}
  +\underbrace{\partial_x(hu)+\partial_y(hv)}_{\text{advection}}
  =\underbrace{S_h}_{\text{source/sink}},
\\[6pt]
\text{(x--momentum)}\;&
  \underbrace{\partial_t(hu)}_{\text{local inertia}}
  +\underbrace{\partial_x\!\bigl(hu^{2}\bigr)+\partial_y(huv)}_{\text{advection}}
  -\underbrace{f\,h\,v}_{\text{Coriolis}}
  +\underbrace{\nu_h\nabla^{2}(hu)}_{\text{horizontal viscosity}}
  +\underbrace{g\,h\,\partial_x\eta}_{\text{gradient}}
\\ &\quad
  =\underbrace{-\,C_f\,u\,\lVert\mathbf{u}\rVert
               +\dfrac{\tau_x}{\rho}}_{\text{friction/forcing}}
   +\underbrace{S_u}_{\text{source/sink}},
\\[6pt]
\text{(y--momentum)}\;&
  \underbrace{\partial_t(hv)}_{\text{local inertia}}
  +\underbrace{\partial_x(huv)+\partial_y(hv^{2})}_{\text{advection}}
  +\underbrace{f\,h\,u}_{\text{Coriolis}}
  +\underbrace{\nu_h\nabla^{2}(hv)}_{\text{horizontal viscosity}}
  +\underbrace{g\,h\,\partial_y\eta}_{\text{gradient}}
\\ &\quad
  =\underbrace{-\,C_f\,v\,\lVert\mathbf{u}\rVert
               +\dfrac{\tau_y}{\rho}}_{\text{friction/forcing}}
   +\underbrace{S_v}_{\text{source/sink}}.
\end{aligned}
$%
\end{adjustbox}
\label{eq:SWE}
\end{equation}

where $h = \eta - B$ represents the total water depth, $\eta$ denotes the free-surface elevation relative to a fixed datum, $B$ is the bed elevation; $\mathbf{u} = (u, v)^\top$ is the depth-averaged velocity vector and $\|\mathbf{u}\| = \sqrt{u^2 + v^2}$; $g$ is gravitational acceleration; $\rho$ represents water density; $\nu_h$ denotes horizontal eddy-viscosity coefficient; $\tau_x$ and $\tau_y$ represent wind stresses in the $x$- and $y$-directions; $f$ is Coriolis parameter; $C_f$ denotes bed friction coefficient; $S_h$, $S_u$, and $S_v$ are generic source/sink terms; and $\nabla^2 (\cdot) = \partial_{xx}(\cdot) + \partial_{yy}(\cdot)$. The depth-averaged formulation captures the essential physics of flood propagation while maintaining computational efficiency necessary for flood simulations.

In principle, the shallow-water system couples a continuity equation, which enforces conservation of mass within the control volume, with two momentum equations that implement Newton’s second law of motion, $F = ma$, in the horizontal directions \citep{Kampf2009}. The momentum balances state that the local plus advective change in momentum equals the sum of all forces acting on the fluid parcel, including pressure gradients, gravitational forces, bed friction, wind stress, Coriolis acceleration, and any other prescribed source or sink terms \citep{Stewart2008}.

For regional flood dynamics simulations, the classic SWE can be further simplified as SSWE, which is employed as the fundamental governing physics for both the hydrodynamic modeling and deep learning components in this study. The SSWE retains the essential physics of Newton's laws while being derived from the incompressible Navier–Stokes equations through depth-integration under the hydrostatic pressure assumption. This formulation is particularly well-suited for coastal and estuarine environments where horizontal length scales significantly exceed vertical dimensions, making the hydrostatic approximation valid for the physics governing CF dynamics. The consistent use of SSWE across both the physics-based SFINCS model and the physics-informed neural network ensures that both approaches are grounded in the same fundamental physical principles. For a flow domain with horizontal coordinates $x, y$ and time $t$, the state vector is defined as:

\begin{equation}
\mathbf{U} = 
\begin{bmatrix}
h \\
\mathbf{q}
\end{bmatrix}^\top
\label{eq:statevector}
\end{equation}

where $\mathbf{q} = h \mathbf{u}$ is the unit-width discharge vector.

Neglecting horizontal viscosity effects, which is justified for the relatively coarse spatial resolutions typically employed in regional flood modeling, the governing equations are expressed in conservative form as:

\begin{equation}
\partial_t \mathbf{U} + \nabla \cdot \mathbf{F(U)} = \mathbf{S(U)}
\label{eq:conservative}
\end{equation}

where $\nabla$ denotes the horizontal divergence operator. The flux tensor combines continuity and momentum transport,

\begin{equation}
\mathbf{F(U)} =
\begin{bmatrix}
\mathbf{q} \\
\frac{\mathbf{q} \otimes \mathbf{q}}{h} + \frac{1}{2}g\eta^2 \mathbf{I}
\end{bmatrix}
\label{eq:flux}
\end{equation}

while the source vector incorporates all non-advective forcing mechanisms relevant to CF,

\begin{equation}
\mathbf{S(U)} =
\begin{bmatrix}
r_p - \nabla \cdot \mathbf{q}_r \\
- g h \nabla B - \frac{\tau_b}{\rho} + \frac{\tau_w}{\rho} + Q_r
\end{bmatrix}
\label{eq:source}
\end{equation}

Here $\otimes$ denotes the tensor product of vectors, $\mathbf{I}$ is the $2{\times}2$ identity matrix, $r_z$ denotes the net vertical water flux (rainfall minus infiltration), and $Q_r$ represents lateral discharge inputs from gauged river inflows (point source discharges) or outfall effluent. Neglecting horizontal viscosity terms helps the system maintain its strictly hyperbolic character. This approach captures the essential wave propagation characteristics necessary for CF modeling while avoiding computational complexity and potential numerical stiffness associated with viscous terms. Should \textit{turbulent} mixing or small-scale viscous effects become important for specific applications, diffusive contributions can be readily incorporated as additional flux divergence terms without altering the fundamental balance-law structure of the equations. This flexibility allows the modeling framework to be adapted for different scales and physical processes while maintaining computational efficiency.

Bed friction parameterization follows Manning's empirical formulation, which has been extensively validated for shallow water flows in coastal and riverine environments:

\begin{equation}
\tau_b = \rho g n^2 \frac{\|\mathbf{u}\| \mathbf{u}}{h^{1/3}}
\label{eq:bedfriction}
\end{equation}

where $n$ is the spatially varying Manning's roughness coefficient. The spatial variability of roughness coefficients allows the model to account for different land cover types and bathymetric features that influence flow resistance during a CF event.

Wind stress forcing at 10-meter height is incorporated through the quadratic drag law:

\begin{equation}
\tau_w = \rho_a C_d \|\mathbf{U}_{10}\| \mathbf{U}_{10}
\label{eq:windstress}
\end{equation}

where $\rho_a$ represents the air density, $C_d$ is the standard drag coefficient, and $\mathbf{U}_{10}$ the measured wind velocity vector. Wind forcing becomes particularly important during tropical cyclones when sustained winds can significantly enhance storm surge propagation and coastal water levels.

As outlined above, the conservative formulation of Eq.~\eqref{eq:conservative} simultaneously encodes the fundamental physics of mass conservation and momentum balance within a unified framework. This formulation provides the theoretical foundation linking the physics-based hydrodynamic model to both the forward numerical solver and the PINN surrogate developed in subsequent sections, ensuring consistency in the governing physics across different modeling approaches.

\subsection{ALPINE model}

\begin{figure}[htbp]
    \centering
    \includegraphics[width=\textwidth]{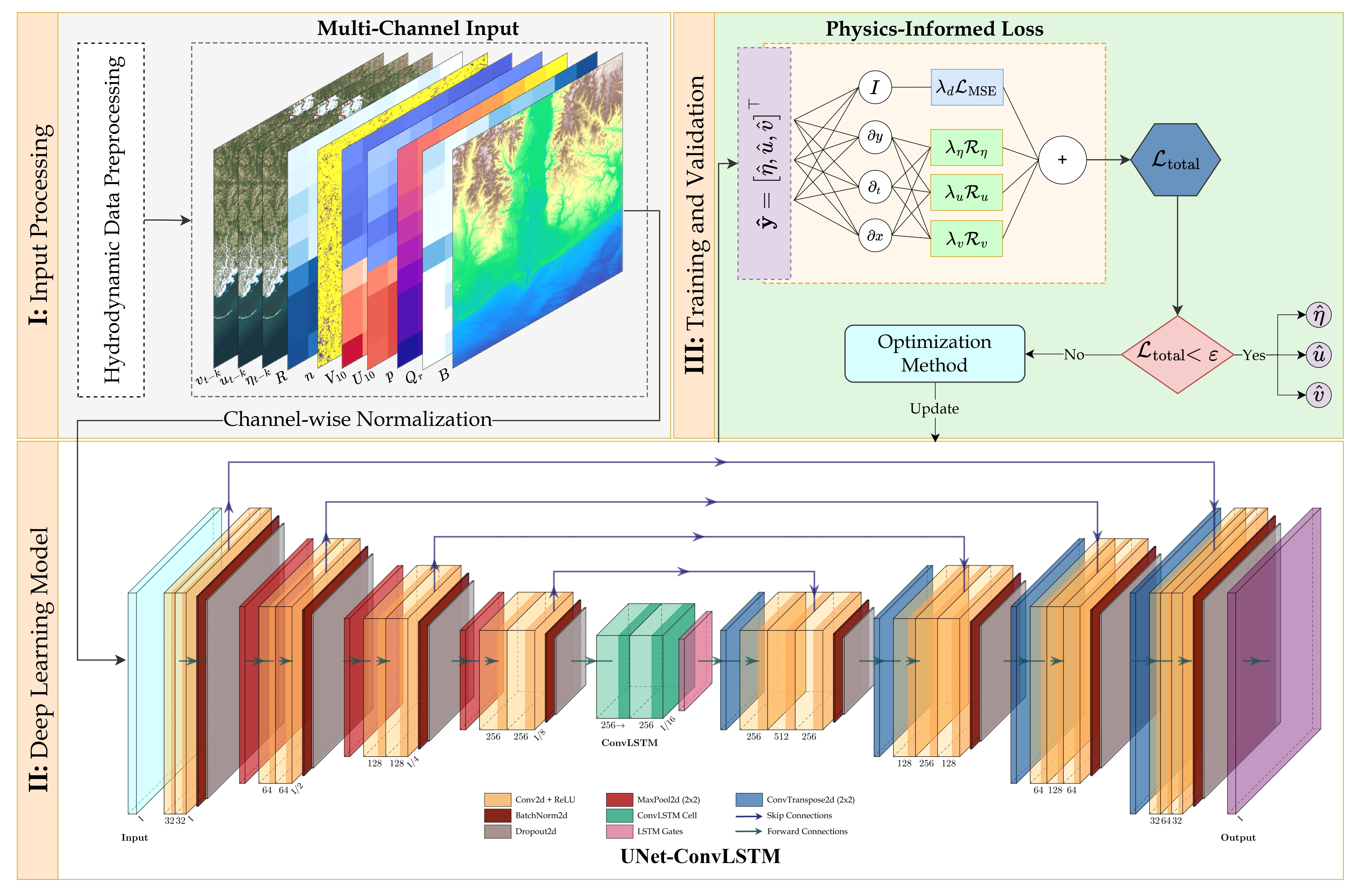}
    \caption{ALPINE model architecture for CF prediction. Schematic representation of the Physics-Informed Neural Network architecture showing the temporal processing component (3D convolution), spatial feature extraction blocks, and the integration of physical constraints through the loss function.
    }
    \label{fig:alpine}
\end{figure}

To emulate shallow-water dynamics under CF conditions, we construct ALPINE that receives a 19-channel input tensor and predicts the water surface elevation $\hat{\eta}_t$ and depth-averaged velocity field $(\hat{u}_t, \hat{v}_t)$ one hour ahead (Figure~\ref{fig:alpine}). The model receives a sequence of spatially distributed inputs, defined over a fixed domain $\Omega \subset \mathbb{R}^2$ discretized into a $536{\times}608$ grid, and is trained using a composite loss that enforces both data fidelity and consistency with the governing equations described in Section~2.3. ALPINE ingests a sequence of five timesteps, $\{t{-}4,\dots,t\}$, so that four previous states supply temporal context during both training and inference. At each time $t$, the input tensor $\mathbf{X}_t$ therefore consists of seven static or forcing fields (i.e., bed elevation $B(x,y)$, Manning’s roughness $n(x,y)$, discharge inflow $Q_r(x,y,t)$, wind forcing $(U_{10}, V_{10})$, rainfall $R(x,y,t)$, sea-level pressure $p(x,y,t)$), and twelve hydrodynamic channels $\left(\eta_{t-k}, u_{t-k}, v_{t-k}\right)_{k=1}^{4}$ that summarise the antecedent flow history.

These inputs are passed through a convolutional encoder--decoder network equipped with a ConvLSTM bottleneck $\mathcal{F}_{\theta}$, yielding the prediction $\hat{\mathbf{y}}_t$:

\begin{equation}
    \hat{\mathbf{y}}_t = \mathcal{F}_{\theta}(\mathbf{X}_t) = [\hat{\eta}_t, \hat{u}_t, \hat{v}_t]^\top
    \tag{8}
\end{equation}

The surrogate consists of a depth-wise temporal convolution with kernel $3{\times}1{\times}1$ couples that now couples five adjacent slices $(t{-}2,t{-}1,t,t{+}1,t{+}2)$, preserving causal structure while enriching short-term dynamics. Each down-sampling stage halves spatial resolution via $2{\times}2$ max pooling and doubles the channel width through a pair of $3{\times}3$ convolution--batch norm--ReLU blocks, followed by dropout. At the bottleneck, temporal memory is injected by a single-step ConvLSTM cell:
\begin{equation}
    (\mathbf{h}_t, \mathbf{c}_t) = \mathcal{C}(\mathbf{e}_t, \mathbf{h}_{t-1}, \mathbf{c}_{t-1})
    \tag{9}
\end{equation}
where $\mathbf{e}_t$ is the encoded feature map at time $t$, and $(\mathbf{h}_t, \mathbf{c}_t)$ denote hidden and cell states. The hidden width is chosen as a multiple $m \in \{1,2,4\}$ of the base channel size. The decoder mirrors the encoder through learned up-sampling and skip connections, terminating in a three-channel output layer.

\begin{figure}[htbp]
    \centering
    \includegraphics[width=\textwidth]{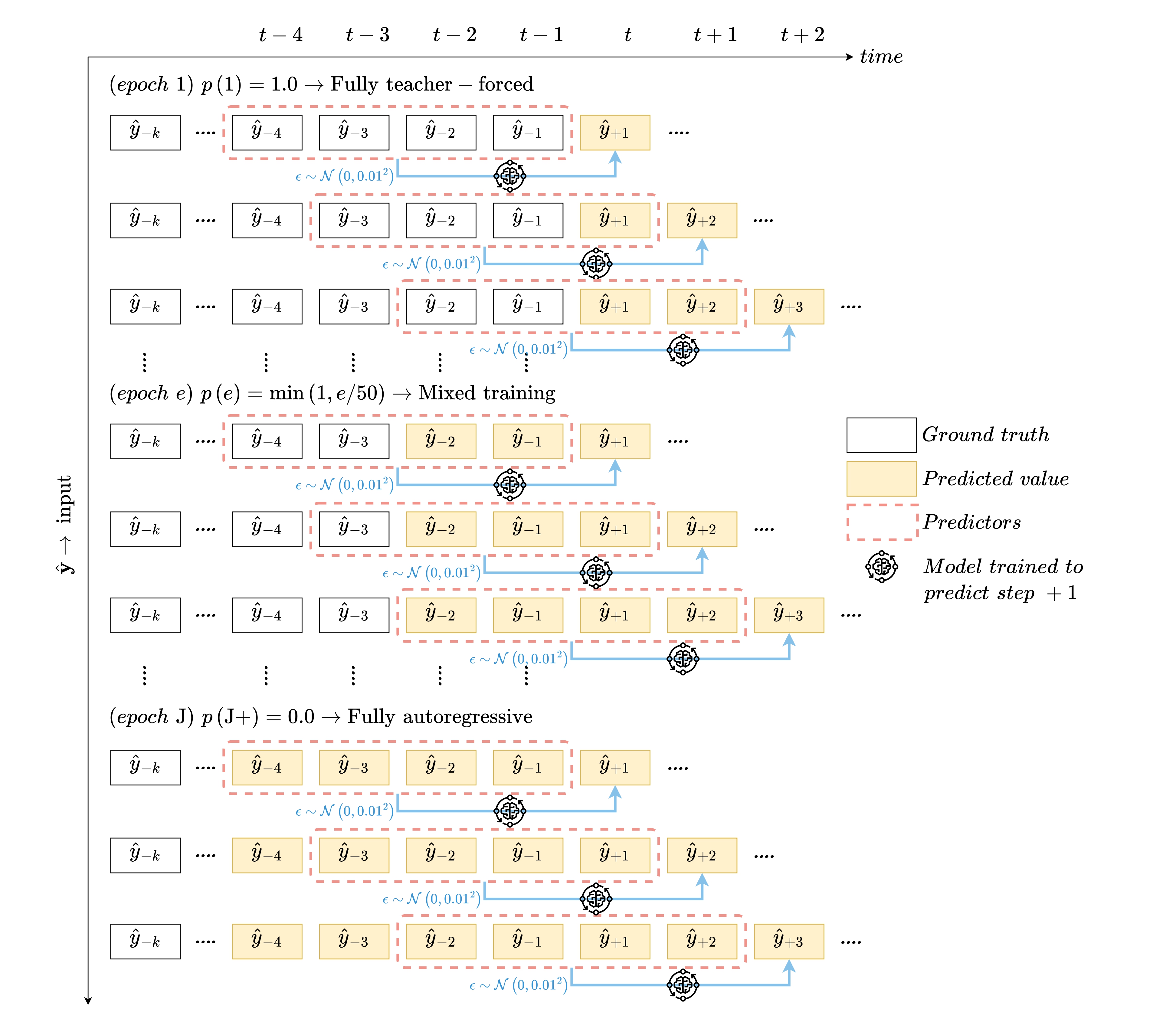}
    \caption{Schematic of an autoregressive rollout with scheduled sampling. At every step the predictor window (red dashed box) is filled with either ground-truth states (white) or previous model outputs (yellow) according to a probability   that decays over epochs. Before the window is fed to the network a single Gaussian perturbation  is added (during training only).}
    \label{fig:autoregres}
\end{figure}

During training we employ an autoregressive roll-out of length eight, i.e. gradients are accumulated through the chain $\{t{-}4,\dots,t{+}3\}$ before a single optimiser step is taken (Figure~\ref{fig:autoregres}). The hidden state of the ConvLSTM is carried forward across the entire roll-out so that temporal dependencies flow through real back-propagation-through-time rather than via truncated gradients. At the first epoch the network is fully teacher-forced: the four history slices embedded in $\mathbf{X}_{t+1}$ are still the SFINCS predictions as ground-truth states $(\eta_{t-3},...,v_t)$. Thereafter we follow a scheduled-sampling scheme in which this replacement occurs with probability $1-p(e)$, where $p(e) = \min(1,e/50)$ increases linearly from zero to unity over the first fifty epochs. Consequently, the model is exposed to its own outputs after only one training iteration, but the proportion of self-feedback grows smoothly, letting the network adapt gradually to the recursive regime.

Immediately after the scheduled sampling selection and before the forward pass, we add a small perturbation $\epsilon \sim \mathcal{N}(0,\,0.01^2)$ to every history channel $\eta_{t-k}, u_{t-k}, v_{t-k}$. This Gaussian noise, applied only during training, forces the model to remain stable when its inputs deviate slightly from the true state and therefore mitigates error amplification over long rollouts. Once a predicted triplet $(\hat{\eta}_{t+1}, \hat{u}_{t+1}, \hat{v}_{t+1})$ has entered the history buffer it is treated as immutable for the remainder of the roll-out, ensuring that length-eight back-propagation faithfully penalises multi-step drift. At inference time no ground-truth fields are ever re-inserted; the surrogate therefore runs in open-loop for the full hurricane sequence.

To further guide the training of $\mathcal{F}_\theta$, we define a composite loss function composed of data fidelity and physics-informed residuals:

\begin{equation}
\mathcal{L}(\theta) = \lambda_d \mathcal{L}_{\text{MSE}} + \lambda_\eta \mathcal{R}_\eta + \lambda_u \mathcal{R}_u + \lambda_v \mathcal{R}_v
\label{eq:loss_total}
\end{equation}

The term $\mathcal{L}_{\text{MSE}}$ denotes the spatially averaged mean squared error:

\begin{equation}
\mathcal{L}_{\text{MSE}} = \frac{1}{|\Omega|} \sum_{(x,y) \in \Omega} \left\| \hat{\mathbf{y}}_t(x,y) - \mathbf{y}_t(x,y) \right\|^2
\label{eq:loss_mse}
\end{equation}

The residuals $\mathcal{R}_\eta$, $\mathcal{R}_u$, and $\mathcal{R}_v$ correspond to the continuity and momentum equations on a three-frame stencil $(t-1, t, t+1)$. To prevent any spurious creation or loss of water, the domain-integrated volume predicted at time $t$ must lie between the volume stored at the previous step plus the ensuing inflow, and the volume observed at the next step minus the intervening outflow. This requirement is imposed through a pair of one-sided forward and backward volume-balance inequalities:

\begin{equation}
\left\{
\begin{aligned}
\mathcal{V}_t(\hat{h}) &\leq \mathcal{V}_{t-1}(\hat{h}) + \Delta t \cdot S_{t-1} \\
\mathcal{V}_{t+1}(\hat{h}) &\leq \mathcal{V}_t(\hat{h}) + \Delta t \cdot S_t
\end{aligned}
\right.
\label{eq:volume_bounds}
\end{equation}

where $\mathcal{V}_t(h) = \iint_\Omega h(x,y,t) \, dx \, dy$ is the total water volume at time $t$, and $S_t = \iint_\Omega \left(R(x,y,t) + Q_r(x,y,t)\right) dx \, dy$ is the net areal source, combining rainfall intensity $R$ and lateral inflow $Q_r$ over the domain $\Omega$. Violation of Eq.~\eqref{eq:volume_bounds} are discouraged through a squared-ReLU penalty:

\begin{equation}
\mathcal{R}_\eta = \left\langle 
\text{ReLU} \left( \frac{\mathcal{V}_t - (\mathcal{V}_{t-1} + \Delta t \cdot S_{t-1})}{|\Omega|} \right)^2 + 
\text{ReLU} \left( \frac{\mathcal{V}_{t+1} - (\mathcal{V}_t + \Delta t \cdot S_t)}{|\Omega|} \right)^2 
\right\rangle
\label{eq:loss_volume}
\end{equation}

where $\text{ReLU}(x) = \max(x, 0)$, and $\langle \cdot \rangle$ denotes the average over all training samples in a batch. This loss term preserves integral mass balance while avoiding the stiffness associated with pointwise PDE residuals—an approach that has recently proven effective in physics-informed flood surrogate models \citep{Donnelly2024, Taghizadeh2025}.

Momentum conservation is enforced locally through residuals derived from the conservative SSWE. For each component $j \in \{x, y\}$, the residual is expressed as:

\begin{equation}
r^{u_j}(x,y) = h \, \partial_t u_j + \nabla \cdot (h u_j \mathbf{u}) + g h \, \partial_j B + \frac{\tau_{b,j}}{\rho} - \frac{\tau_{w,j}}{\rho}
\label{eq:residual_momentum}
\end{equation}

where $\mathbf{u} = (u, v)$ is the depth-averaged velocity. Shear stresses $\tau_{b,j}$ and $\tau_{w,j}$ are computed from the friction and wind models described in Section~2.3. The momentum loss terms are then defined as:

\begin{equation}
\mathcal{R}_{u_j} = \left\langle (r^{u_j})^2 \right\rangle
\label{eq:loss_momentum}
\end{equation}

and are included in the total loss (Eq.~\eqref{eq:loss_total}) as $\mathcal{R}_u = \mathcal{R}_{u_x}$ and $\mathcal{R}_v = \mathcal{R}_{u_y}$.

The surrogate is trained using five historical flood events—Hurricane Isaac (2012), Hurricane Nate (2017), Hurricane Michael (2018), Tropical Storm Claudette (2021), and Hurricane Ida (2021). Validation is performed on held-out sequences from Ida (2021), while Hurricane Francine (2024) served as an unseen test case for blind evaluation. Log-scaling, min-max scaling, and z-scoring are used to normalize inputs and outputs based on distribution characteristics in order to address the different scale among features.

To optimize the network architecture, an extensive hyperparameter search is conducted to determine the optimal learning rate and relative weights $(\lambda_d, \lambda_\eta, \lambda_u, \lambda_v)$ ensuring that no single term dominates the loss landscape. Additional configurations of $\mathcal{F}_\theta$ explored include variations in base channel widths (16, 32, 64), network depth levels (2, 3, 4), ConvLSTM bottleneck widths scaled by factors of 1, 2, and 4, kernel sizes (3 and 5), and dropout probabilities (0.0, 0.1, 0.2). We ran 50 trials across this hyperparameter grid and selected the configuration that minimized validation loss. The chosen model was then retrained for up to 200 epochs using the AdamW optimizer. Two training callbacks were used to monitor convergence: a checkpoint mechanism that saved weights whenever the validation error decreased, and an early stopping criterion that terminated training if no improvement was seen for 10 consecutive epochs. These measures ensured that the model retained the best-performing configuration while avoiding overfitting.

\section{Results and discussion}
\subsection{SFINCS model}
Figs. S1-S6 presents the track and the spatial distribution of simulated peak water depths across the six storm events. Generally, flood patterns north of the bay vary little among the storms because backwater in that reach is governed primarily by river discharge, whereas differences elsewhere reflect how far surge overtops the open-bay shore and penetrates the upriver. Hurricanes Isaac (Fig. S1) and Ida (Fig. S5) generate the most widespread inundation. Both storms drive 1.5–3 m of water up the Mobile and Tensaw rivers, but Isaac floods a broader section of the western shoreline than Ida. The slow movement of Isaac resulted in prolonged wind, coastal flooding and its heavy flooding rains, especially over southeast Mississippi, also impacted southwest Alabama, leading to inland flooding \citep{Berg2013}. Hurricane Francine (Fig. S6) produces a similar inland response to Hurricane Isaac. Its peak depths of 1–3 m extend across much of the bay interior and the upper floodplain, while surge was minor to moderate, remained mostly below 1 m along the open coast \citep{Bucci2025}. Hurricane Nate (Fig. S2) raises 0.5–1 m of water on Dauphin Island and the Fort Morgan peninsula; inside the delta, levels fall below 1.5 m, and the upper floodplain stays largely dry. This event was primarily a storm surge event, leading to 0.9–1.8 m of storm surge inundation across coastal Baldwin County, on the eastern shore of Mobile Bay \citep{Beven2018}. 

Tropical Strom Claudette (Fig. S4) confines water to the main distributary channels, with depths generally under 2 m and little overbank flow beyond the barrier islands. Claudette produced substantial rainfall that resulted in flash flooding across portions of the Gulf Coast \citep{Papin2022}. Hurricane Michael (Fig. S3) shows the smallest overall signal as it made landfall in the Florida Panhandle, far to the east of Mobile Bay \citep{Hagen2019}. Hurricane impacts are typically most severe in the right-front quadrant relative to the storm's forward motion, which placed Mobile Bay in a less affected zone during Michael's passage \citep{Elsner1999}. The coastal run-up due to Hurricane Michael was under 0.5 m and the delta depths seldom exceed 1 m. These varying patterns highlight the complex nature of CF, where storm track, timing of peak surge relative to high tide, antecedent rainfall, and river discharge conditions can interact at various levels and lead to nuanced differences in flooding impacts.

The SFINCS model's performance was evaluated using water level data from five NOAA tide-gauge stations: Dog River Bridge (Station ID: 8735391), East Fowl River Bridge (8735523), Dauphin Island (8735180), Coast Guard Sector (8736897), and Mobile State Docks (8737048). The evaluation metrics included root-mean squared error (RMSE), Nash--Sutcliffe efficiency (NSE, \citealp{Nash1970}), and coefficient of determination ($\mathrm{R}^2$). RMSE ranges from 0 to positive infinity, with values closer to 0 indicating better model performance. NSE ranges from negative infinity to 1, with a value of 1 indicating perfect agreement between observations and model simulations. Additionally, $\mathrm{R}^2$ is used to evaluate the proportion of variance in observed data explained by the model predictions. $\mathrm{R}^2$ ranges from 0 to 1, where values closer to 1 indicate that a larger proportion of the variance is explained by the model, with a value of 1 indicating perfect linear relationship between observed and predicted values. The mathematical expressions for these metrics are as follows:

\begin{gather}
  \mathrm{RMSE} = \sqrt{\frac{1}{N}\sum_{i=1}^{N}(O_i - P_i)^2} \tag{16} \label{eq:rmse}\\[6pt]
  \mathrm{NSE} = 1 - \frac{\sum_{i=1}^{N}(O_i - P_i)^2}{\sum_{i=1}^{N}(O_i - \overline{O})^2} \tag{17} \label{eq:nse}\\[6pt]
  \mathrm{R}^2 = \left( \frac{\sum_{i=1}^{N}(O_i - \overline{O})(P_i - \overline{P})}%
              {\sqrt{\sum_{i=1}^{N}(O_i - \overline{O})^2 \sum_{i=1}^{N}(P_i - \overline{P})^2}} \right)^2
  \tag{18} \label{eq:r2}
\end{gather}

where $O_i$ and $P_i$ denote observed and predicted values, $\overline{O}$ and $\overline{P}$ are their means, and $N$ is the sample size.

\begin{figure}[htbp]
    \centering
    \includegraphics[width=\textwidth]{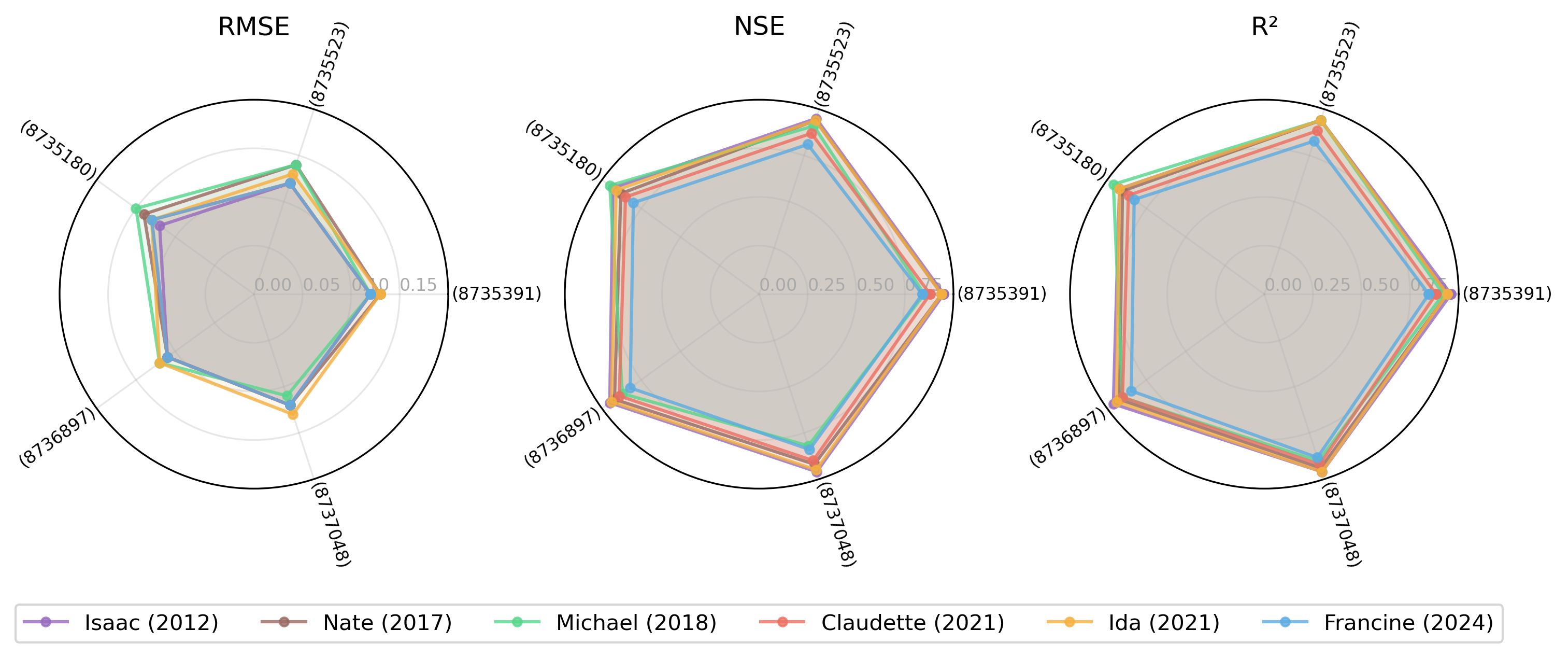}
    \caption{SFINCS model performance evaluation across six storm events using radar plots of three key metrics: Root Mean Square Error (RMSE), Nash-Sutcliffe Efficiency (NSE), and coefficient of determination (R²). Each plot displays model performance at five NOAA tide-gauge stations: Dog River Bridge (Station ID: 8735391; 30.57° N, 88.09° W), East Fowl River Bridge (Station ID: 8735523; 30.44° N, 88.11° W), Dauphin Island (Station ID: 8735180; 30.250° N, 88.075° W), Coast Guard Sector Mobile (Station ID: 8736897; 30.65° N, 88.06° W), and Mobile State Docks (Station ID: 8737048; 30.71° N, 88.04° W). The colored lines represent different storm events: Isaac (2012) in purple, Nate (2017) in brown, Michael (2018) in green, Claudette (2021) in red, Ida (2021) in orange, and Francine (2024) in blue.}
    \label{fig:sfincs}
\end{figure}

As a rough estimate, for hurricane storm surge modeling along the U.S. Gulf-Atlantic Coast, RMSE values below 0.20 m are considered acceptable for simulating extreme events \citep{riverside2015}. Overall, the model showed good agreement with observations across all evaluated storm events and stations (\ref{fig:sfincs} and Table S1). RMSE values ranged from 0.05 to 0.09 m across all stations and events, well below the 0.20 m threshold for acceptable performance. The lowest RMSE values (0.05-0.06 m) were generally observed for Hurricane Michael (2018), while slightly higher values occurred during Hurricane Isaac (2012). NSE values consistently exceeded 0.80 across all scenarios, with most values above 0.90, indicating excellent model skill in capturing temporal dynamics. Similarly, $\mathrm{R}^2$ values ranged from 0.83 to 0.96, demonstrating that the model explains 83-96\% of the variance in observed water levels across all events and stations.

These results indicate that the model successfully captured both the timing and magnitude of water level variations during each simulation period. Peak water levels were particularly well represented during the most intense phases of each event, as evidenced by the high NSE and $\mathrm{R}^2$ values. Station-specific performance showed some variation, with coastal stations (e.g., Dauphin Island) generally exhibiting slightly better performance than upstream riverine locations. This pattern reflects the model's strength in capturing tidal and surge dynamics, though riverine processes are also well represented. The good performance across all stations validates the model's ability to simulate the complex interactions between coastal surge, tidal dynamics, and riverine discharge that characterize CF events in Mobile Bay. These results confirm that the SFINCS model provides a reliable foundation for the subsequent PINN development, ensuring that a verified set of data used for training captures the essential physics of CF dynamics.

\subsection{ALPINE model results}

This section presents the development and validation of the PINN surrogate model for CF prediction in Mobile Bay, comparing its performance against the baseline UNet-ConvLSTM model through comprehensive spatial and temporal evaluation metrics. Table~\ref{tab:best_model_config} details the architectural configurations derived from hyperparameter search for both models. The grid search explored multiple configurations across key architectural components to identify optimal settings for CF prediction. The PINN architecture employs 64 base channels in its convolutional layers, doubling the 32 channels used in the UNet-ConvLSTM baseline. This increased channel capacity allows PINN to learn richer feature representations necessary for encoding both the flood dynamics and physical constraints.

Both architectures share a common UNet depth of 4 layers, which provides sufficient hierarchical feature extraction while maintaining computational efficiency. The encoder progressively downsamples the spatial resolution through max pooling operations, while the decoder reconstructs the full resolution through learned upsampling and skip connections. The models employ $5{\times}5$ convolutional kernels throughout the network, providing larger receptive fields compared to standard $3{\times}3$ kernels. The models utilize an LSTM hidden multiplier of 4.0, quadrupling the base channel dimension at the bottleneck layer. This expanded hidden state dimension enables complex temporal modeling, allowing the networks to better capture the evolution of CF events over the prediction horizon. The UNet-ConvLSTM applies no dropout during training, while ALPINE uses a dropout rate of 0.1 to prevent overfitting, particularly important given the limited number of historical storm events available for training and the additional complexity of physics-informed constraints.

The architectural choices result in substantially different model complexities. ALPINE contains 110,112,259 trainable parameters compared to 27,536,131 for the UNet-ConvLSTM, approximately a 4-fold difference. Increased parameter counts in ALPINE stem primarily from the larger base channels (64 vs 32) and the regularization requirements, which are necessary to simultaneously learn the data-driven flood patterns and satisfy the physics-based constraints imposed through the loss function.

\begin{table}[htbp]
\caption{Details of the best model configurations obtained from hyperparameter search.}
\centering
\begin{tabular}{lcc}
\toprule
\textbf{Component} & \textbf{UNet-ConvLSTM} & \textbf{PINN} \\
\midrule
Base channels           & 32          & 64 \\
UNet depth              & 4           & 4 \\
LSTM hidden multiplier  & 4.0         & 4.0 \\
Kernel size             & 5           & 5 \\
Dropout rate            & 0.0         & 0.1 \\
Total parameters        & 27,536,131  & 110,107,075 \\
\bottomrule
\end{tabular}
\label{tab:best_model_config}
\end{table}

\begin{figure}[htbp]
    \centering
    \includegraphics[width=0.8\textwidth]{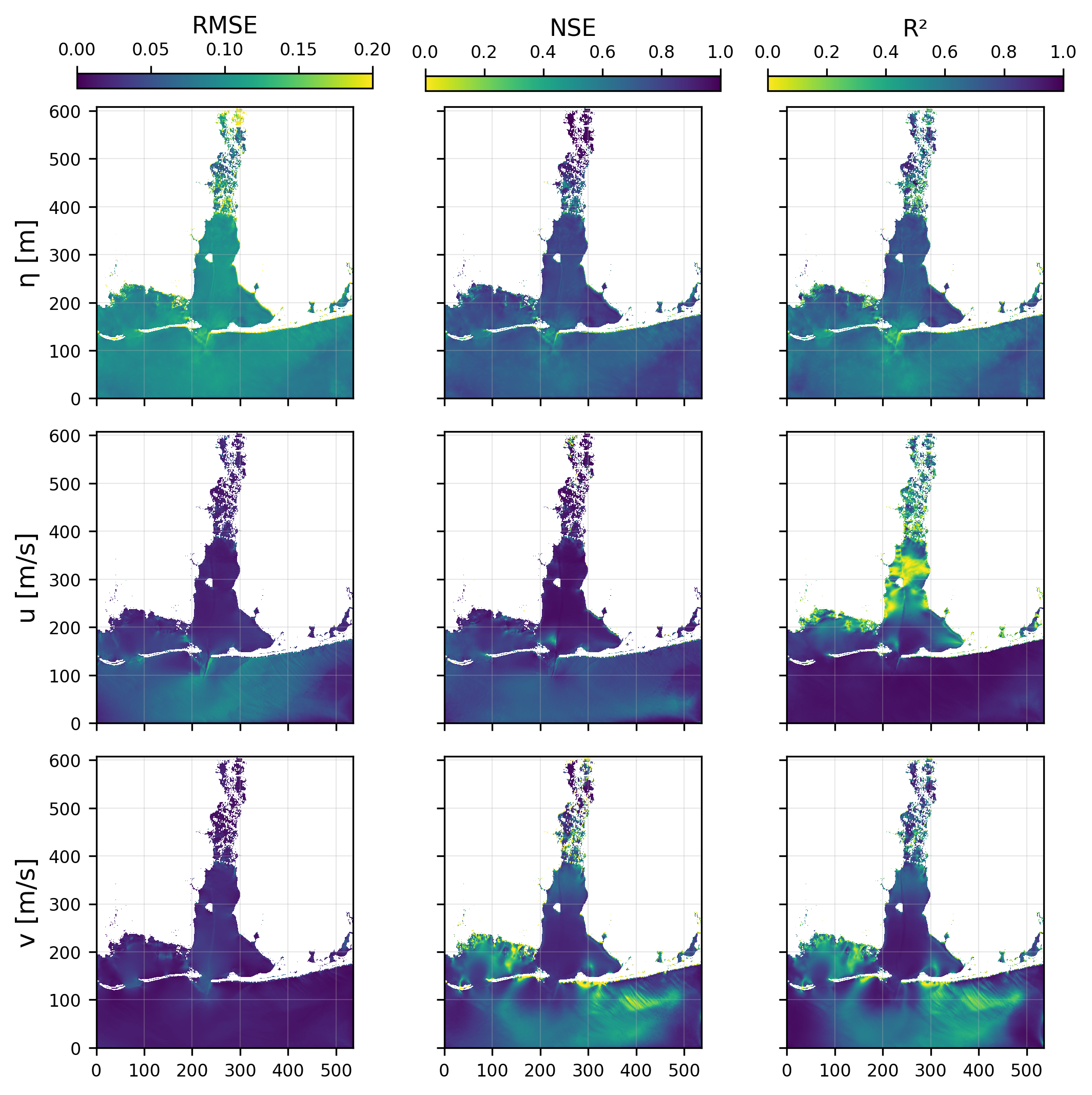}
    \caption{Spatial distribution of UNet-ConvLSTM model performance metrics for water surface elevation ($\eta$), eastward depth-averaged velocity ($u$), and northward depth-averaged velocity ($v$) predictions during Hurricane Francine. The figure displays four key performance metrics computed spatially across the computational domain: Root Mean Square Error (RMSE), Nash-Sutcliffe Efficiency (NSE), and coefficient of determination ($\mathrm{R}^2$). Each row represents a different variable: $\eta$ (top), $u$ (middle), and $v$ (bottom).
    }
    \label{fig:francine_map1}
\end{figure}

\begin{figure}[htbp]
    \centering
    \includegraphics[width=0.8\textwidth]{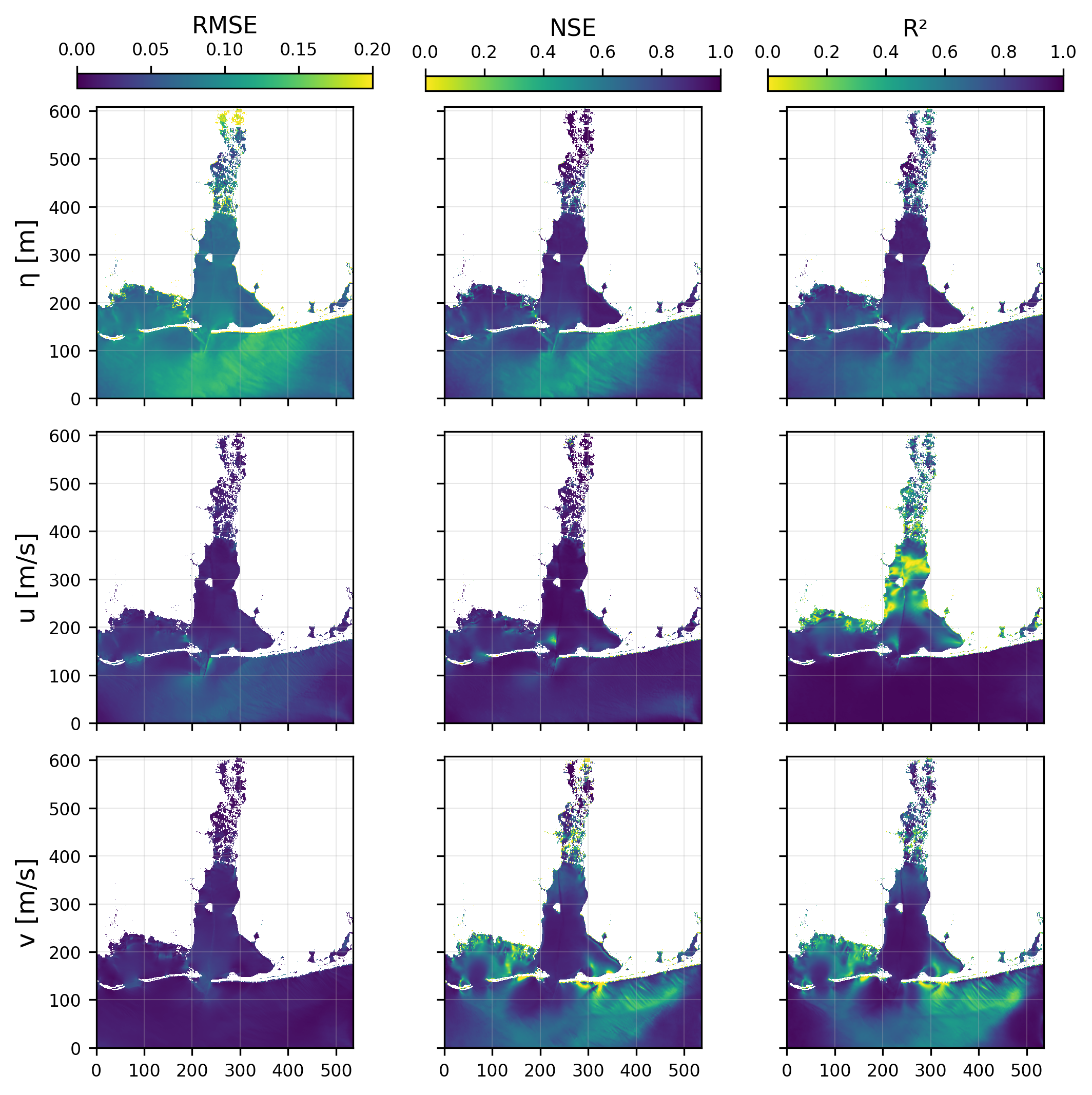}
    \caption{Spatial distribution of ALPINE model performance metrics for water surface elevation ($\eta$), eastward depth-averaged velocity ($u$), and northward depth-averaged velocity ($v$) predictions during Hurricane Francine. The figure displays four key performance metrics computed spatially across the computational domain: Root Mean Square Error (RMSE), Nash-Sutcliffe Efficiency (NSE), and coefficient of determination ($\mathrm{R}^2$). Each row represents a different variable: $\eta$ (top), $u$ (middle), and $v$ (bottom).
    }
    \label{fig:francine_map2}
\end{figure}

The spatial performance of both models during Hurricane Francine, an unseen test event, reveals distinct patterns in prediction accuracy across the Mobile Bay domain. Figure~\ref{fig:francine_map1} and Figure~\ref{fig:francine_map2} present the spatial distribution of four evaluation metrics (Eqs.~\ref{eq:rmse}--\ref{eq:r2}) for water surface elevation ($\eta$), eastward velocity ($u$), and northward velocity ($v$) components.

For $\eta$ predictions, the UNet-ConvLSTM model (Figure~\ref{fig:francine_map1}, top row) exhibits RMSE values ranging from near 0.05~m in the western and central bay waters to $>$0.15~m over the northern delta regions. The elevated errors in these shallow areas reflect the model's difficulty in capturing complex wetting and drying processes during CF, where rapid transitions between exposed mudflats and inundated zones amplify local depth errors. The NSE values for the baseline model show strong performance ($>$0.8) in the open waters, northern parts, and most eastern and western parts of the bay. However, it deteriorates significantly ($<$0.5) in two regions: in the northern riverine sections where the Mobile and Tensaw rivers enter the bay, and near the barrier islands at the bay entrance. These zones experience strong hydraulic gradients and momentum exchanges among river discharge, wind setup, and seaward surge, conditions that magnify any phase lag in the purely data-driven model. The $\mathrm{R}^2$ values show consistent patterns with NSE and strong correlations ($>$0.8) across all regions except the regions with lower NSE.

In contrast, the ALPINE model (Figure~\ref{fig:francine_map2}, top row) demonstrates improved performance across all metrics. The RMSE values remain below 0.10~m throughout most of the domain, with only isolated areas along the extreme northern boundaries exceeding 0.15~m. By enforcing mass and momentum balance, the ALPINE model better reproduces the steep water-surface slopes generated by river inflows meeting wind-driven surge, thereby reducing error hot spots at channel junctions and tidal flats. The NSE values maintain consistently high levels ($>$0.8) across the entire bay, including the challenging northern delta region where the baseline model struggled. This improvement is particularly evident along the eastern shore near the barrier islands and western bays, where the ALPINE model achieves NSE values about 0.8 compared to 0.4--0.6 for the baseline model. This gain shows the value of embedding shallow-water physics to control over-smoothing in zones with sharp bathymetric breaks.

For $u$, both models show the largest RMSE and correspondingly degraded NSE in the western bay and near the barrier islands, where fetch-aligned winds force strong east--west jets that interact with the narrow inlets \citep{Noble1996}. For example, the UNet-ConvLSTM produces RMSE values up to 0.15~m/s in these regions, while the ALPINE model limits errors to below 0.10~m/s by better conserving momentum through the constricted passes. Notably, the highest discrepancy between our data-driven models and the SFINCS simulations appears in the $\mathrm{R}^2$ metric within the narrow tidal channels and river mouths where flow velocities are greatest. This degraded correlation likely stems from the highly dynamic and nonlinear flow patterns in these constricted passages, where small spatial shifts in predicted velocity fields can significantly impact correlation metrics despite reasonable magnitude predictions.

The $v$ predictions present the greatest challenge for both models, with lower overall NSE values compared to the other variables. Both models achieve positive NSE inside Mobile Bay, with negative values indicating poor performance in the shallow western margins and in the Gulf of Mexico outside the bay entrance. Northward flow in these sectors is governed by a balance among pressure-gradient acceleration, Coriolis deflection, and return-flow compensation for strong eastward winds—interactions that are difficult to resolve without explicitly modeling vertical shear and stratification \citep{Coogan2020, Ralston2024}. Consequently, the complex interaction between bay circulation and Gulf waters, where tidal exchanges, wind-driven currents, and density gradients create multidirectional flow patterns, challenges purely surface-based surrogates.

\begin{figure}[htbp]
    \centering
    \includegraphics[width=0.7\textwidth]{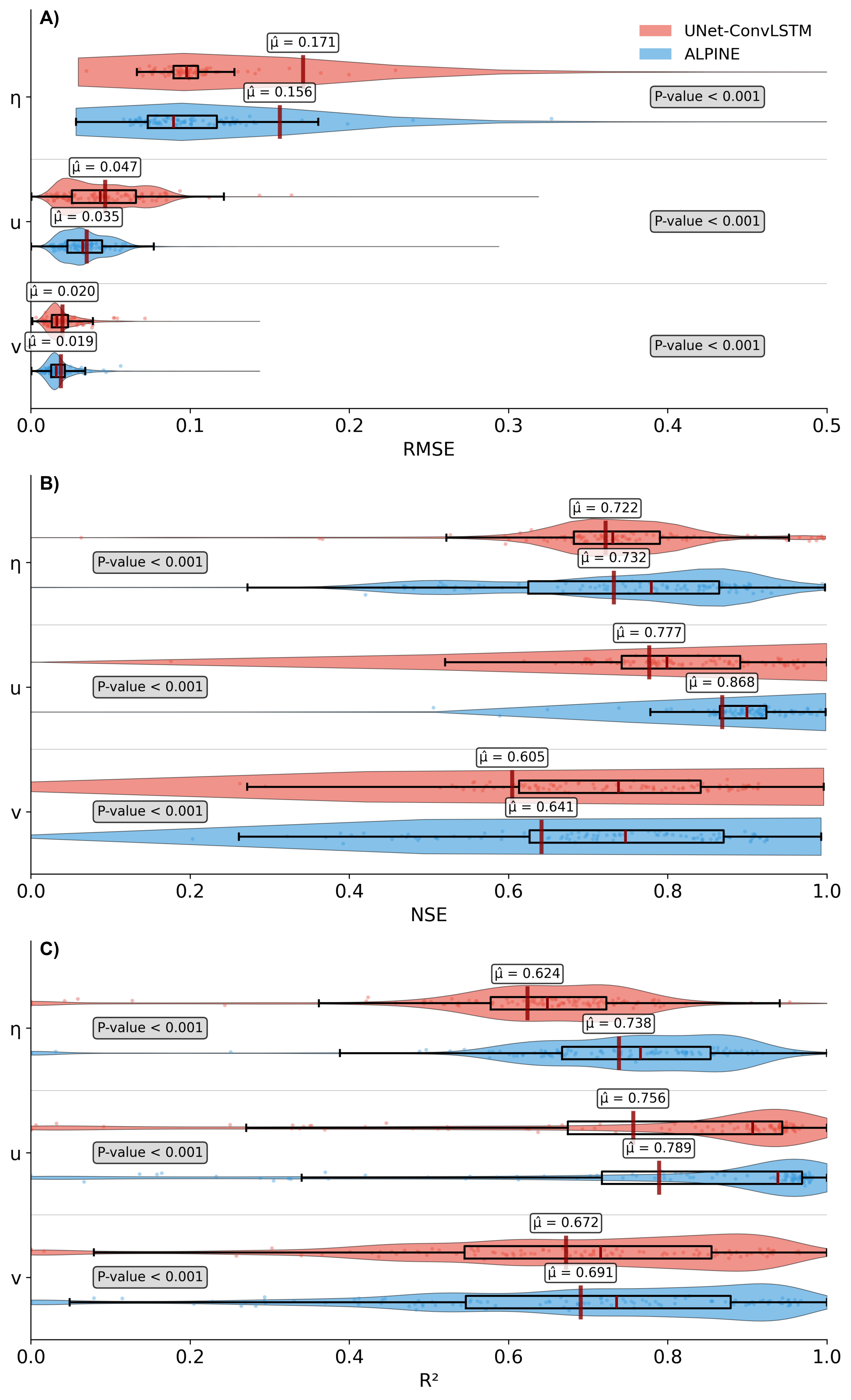}
    \caption{Box-whisker plots comparing pixel-wise performance metrics between UNet-ConvLSTM (red) and ALPINE (blue) models for water surface elevation ($\eta$), eastward velocity ($u$), and northward velocity ($v$) predictions during Hurricane Francine. Panel A shows Root Mean Square Error (RMSE), Panel B displays Nash--Sutcliffe Efficiency (NSE), and Panel C presents coefficient of determination ($\mathrm{R}^2$). Each distribution represents the statistical summary of metric values computed at every pixel across the computational domain, with boxes indicating interquartile ranges, whiskers extending to 1.5 times the interquartile range, and the central line showing the median value. Mean values ($\hat{\mu}$) are annotated for each distribution. P-values from Wilcoxon signed-rank tests indicate statistical significance of performance differences between models (P-value $<$ 0.001 denotes statistical significance). The violin plots illustrate the full probability density distribution of pixel-wise performance metrics across the Mobile Bay domain.
    }
    \label{fig:raincloud}
\end{figure}

Figure~\ref{fig:raincloud} provides a statistical comparison of pixel-wise performance between the UNet-ConvLSTM and ALPINE models across the entire Mobile Bay domain during Hurricane Francine. The box-whisker plots reveal the distribution characteristics of each performance metric computed at every computational grid cell, offering insights into both central tendencies and spatial variability in model accuracy

For RMSE (Figure~\ref{fig:raincloud}A), the ALPINE model demonstrates consistently lower error distributions across all three variables. The $\eta$ predictions show the most substantial improvement, with ALPINE achieving a median RMSE of 0.156~m compared to 0.171~m for UNet-ConvLSTM, representing an 8.8\% reduction in typical prediction errors. The interquartile range (IQR) for $\eta$ is wider in ALPINE than in the baseline, but the physics-informed model eliminates most of the high-error tail beyond 0.30~m. For the velocity components, particularly $u$, ALPINE’s IQR is noticeably narrower, indicating more consistent performance. For velocity components, ALPINE shows particularly strong improvements in $u$ predictions, with median RMSE decreasing from 0.047~m~s$^{-1}$ to 0.035~m~s$^{-1}$. This 25.5\% reduction reflects better momentum conservation in the dominant eastward flow patterns.

The NSE distributions (Figure~\ref{fig:raincloud}B) of ALPINE show improvements in both central tendencies for $\eta$ with notably different distribution shapes. For $\eta$ predictions, both models achieve high median NSE values ($> 0.72$), and ALPINE shifts the distribution toward higher values while removing most of the lowest-skill outliers. The $u$ component shows the largest NSE improvement, with ALPINE’s mean increasing from 0.777 to 0.868, indicating substantially better capture of velocity field dynamics. Conversely, $v$ predictions remain challenging for both models, with median NSE values below 0.7, and ALPINE offers only a modest gain with little change in variability.

The $\mathrm{R}^2$ distributions (Figure~\ref{fig:raincloud}C) demonstrate ALPINE’s enhanced ability to explain variance in observed data. The $\eta$ predictions indicate that the physics-informed model explains about 74\% of water-level variance versus 62\% for the baseline. The $u$ component shows consistent improvement with ALPINE achieving $\mathrm{R}^2$ values above 0.78, while $v$ predictions again present the greatest challenge with $\mathrm{R}^2$ values around 0.67--0.69 for both models. Statistical significance testing using Wilcoxon signed-rank tests confirms that ALPINE’s improvements are statistically significant ($p < 0.001$) across all metrics and variables. The violin-plot shapes reveal that ALPINE not only improves central performance metrics but also reduces the occurrence of extreme outliers, a skill critical for operational flood forecasting, where consistent reliability across the entire domain is essential. Detailed summary statistics for these distributions are provided in Table~\ref{tab:summary_performance}.

\begin{table}[htbp]
\caption{Statistical summary of spatial performance metrics for ALPINE and UNet-ConvLSTM models.}
\centering
\begin{tabular}{clccc}
\toprule
\textbf{Variable} & \textbf{Metric} & \textbf{UNet-ConvLSTM} & \textbf{PINN} & \textbf{Mean improvement (\%)} \\
\midrule
\multirow{3}{*}{$\eta$} 
    & RMSE & $0.171 \pm 0.47$ & $0.156 \pm 0.41$ & 8.77 \\
    & NSE  & $0.722 \pm 0.16$ & $0.732 \pm 0.35$ & 1.39 \\
    & R$^2$  & $0.624 \pm 0.17$ & $0.738 \pm 0.17$ & 18.27 \\
\midrule
\multirow{3}{*}{$u$} 
    & RMSE & $0.047 \pm 0.02$ & $0.035 \pm 0.02$ & 25.53 \\
    & NSE  & $0.777 \pm 1.34$ & $0.868 \pm 1.23$ & 11.71 \\
    & R$^2$  & $0.756 \pm 0.28$ & $0.789 \pm 0.27$ & 4.37 \\
\midrule
\multirow{3}{*}{$v$} 
    & RMSE & $0.020 \pm 0.01$ & $0.019 \pm 0.01$ & 5.00 \\
    & NSE  & $0.605 \pm 0.23$ & $0.641 \pm 0.23$ & 5.95 \\
    & R$^2$  & $0.672 \pm 0.23$ & $0.691 \pm 0.23$ & 2.83 \\
\bottomrule
\end{tabular}
\label{tab:summary_performance}
\end{table}

\begin{figure}[htbp]
    \centering
    \includegraphics[width=\textwidth]{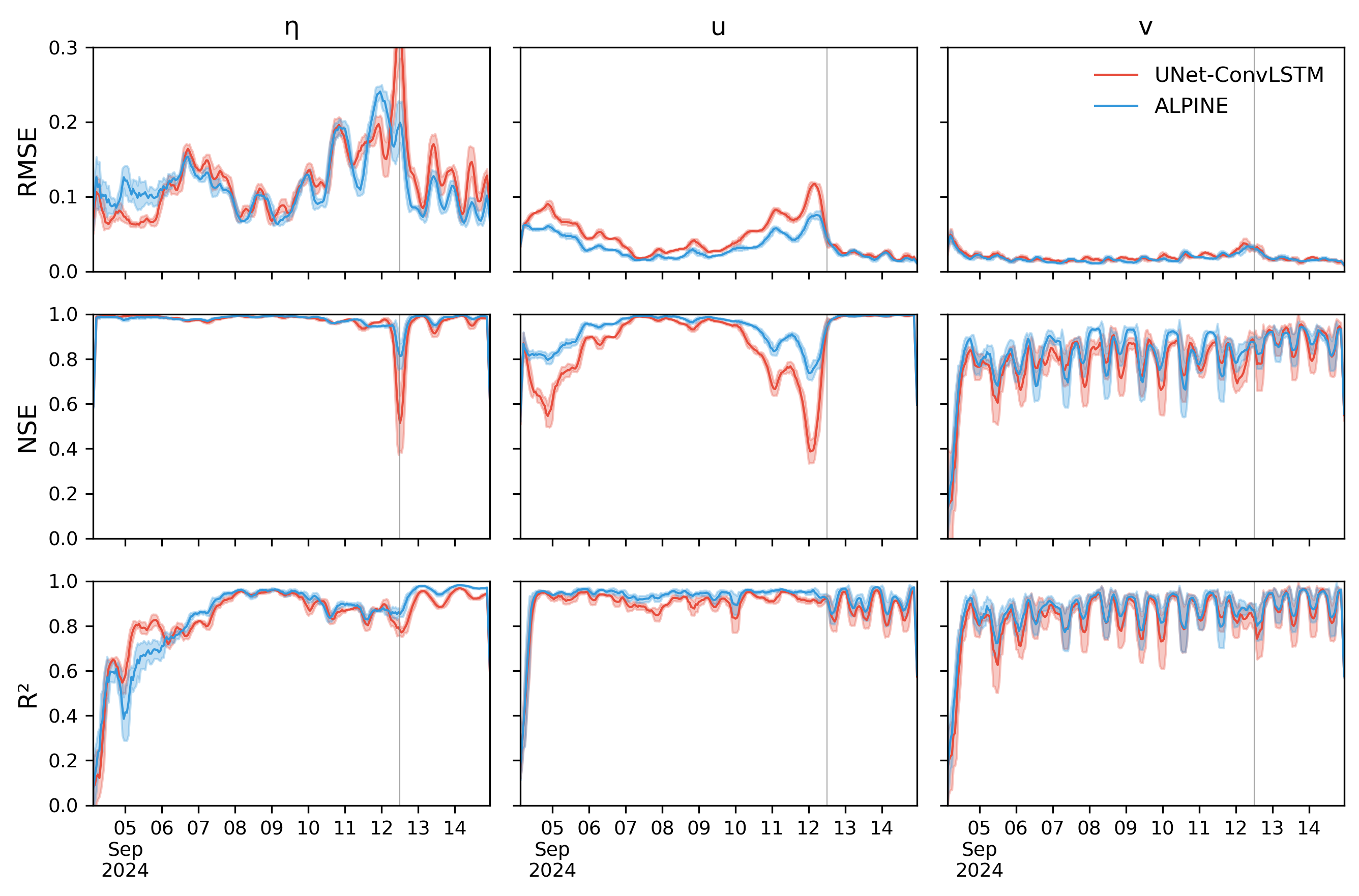}
    \caption{Temporal evolution of domain-averaged performance metrics for UNet-ConvLSTM (red) and ALPINE (blue) models during Hurricane Francine (September 4-14, 2024). The vertical black line in subplots represents the peak of storm inside the study area. The figure displays Root Mean Square Error (RMSE, top row), Nash-Sutcliffe Efficiency (NSE, middle row), and coefficient of determination ($\mathrm{R}^2$, bottom row) for water surface elevation ($\eta$, left column), eastward velocity ($u$, middle column), and northward velocity ($v$, right column). Each metric represents the spatial average across all computational grid cells at hourly intervals throughout the storm event.
    }
    \label{fig:temporal_metrics}
\end{figure}

Figure~\ref{fig:temporal_metrics} shows the temporal evolution of domain-averaged performance metrics for both models throughout Hurricane Francine (September 4–14, 2024). The hourly time series reveal how model accuracy varies with storm intensity and hydrodynamic conditions.

For $\eta$ predictions (Figure~\ref{fig:temporal_metrics}, left column), both models show relatively stable RMSE values of 0.05--0.15~m during the early storm phases (September 4--7), when surge levels remain modest and riverine discharge dominates the hydrodynamics. However, as the storm intensifies and approaches Mobile Bay (September 8--11), RMSE values begin to diverge noticeably. The UNet-ConvLSTM model exhibits sharp deterioration during peak storm conditions (September 11--13), with RMSE exceeding 0.30~m, while ALPINE maintains substantially lower errors, typically below 0.20~m even during the most intense phases. This performance gap is particularly pronounced during the storm's peak intensity on September 12, when multiple flood drivers interact simultaneously and the purely data-driven model struggles to maintain physical consistency.

The NSE temporal patterns mirror the RMSE trends but provide additional insight into model skill degradation. Both models maintain high NSE values ($>0.9$) during quiescent periods, but the UNet-ConvLSTM shows dramatic performance collapse during peak surge conditions, with NSE dropping below 0.4 on September 12. ALPINE demonstrates remarkable resilience, maintaining NSE values above 0.8 throughout most of the event, with only brief periods of degraded performance. The $\mathrm{R}^2$ evolution shows similar patterns, with ALPINE consistently explaining a higher proportion of water level variance, particularly during the critical peak surge period when accurate predictions are most crucial for emergency management.

For $u$ velocity predictions (Figure~\ref{fig:temporal_metrics}, middle column), the temporal analysis reveals more consistent performance differences between the models. ALPINE maintains lower RMSE values throughout the entire event, with the advantage becoming most pronounced during periods of strong wind-driven currents (September 9--13). The NSE patterns show ALPINE's superior ability to capture velocity field dynamics, with particularly notable improvements during the storm's approach and peak phases when fetch-aligned winds generate strong eastward jets through the bay. Both models show periodic oscillations in performance that correlate with tidal cycles, but ALPINE's physics-informed momentum conservation helps maintain more stable accuracy across these natural variations.

The $v$ velocity component presents the most challenging prediction task for both models, with generally lower NSE and $\mathrm{R}^2$ values in most timesteps. However, ALPINE shows modest but consistent improvements, particularly during the storm's intensification phase. The temporal variability in $v$ predictions reflects the complex three-dimensional circulation patterns that develop as storm surge interacts with riverine inflows and wind stress, processes that challenge the depth-averaged modeling framework employed by both surrogates.

\begin{figure}[htbp]
    \centering
    \includegraphics[width=0.7\textwidth]{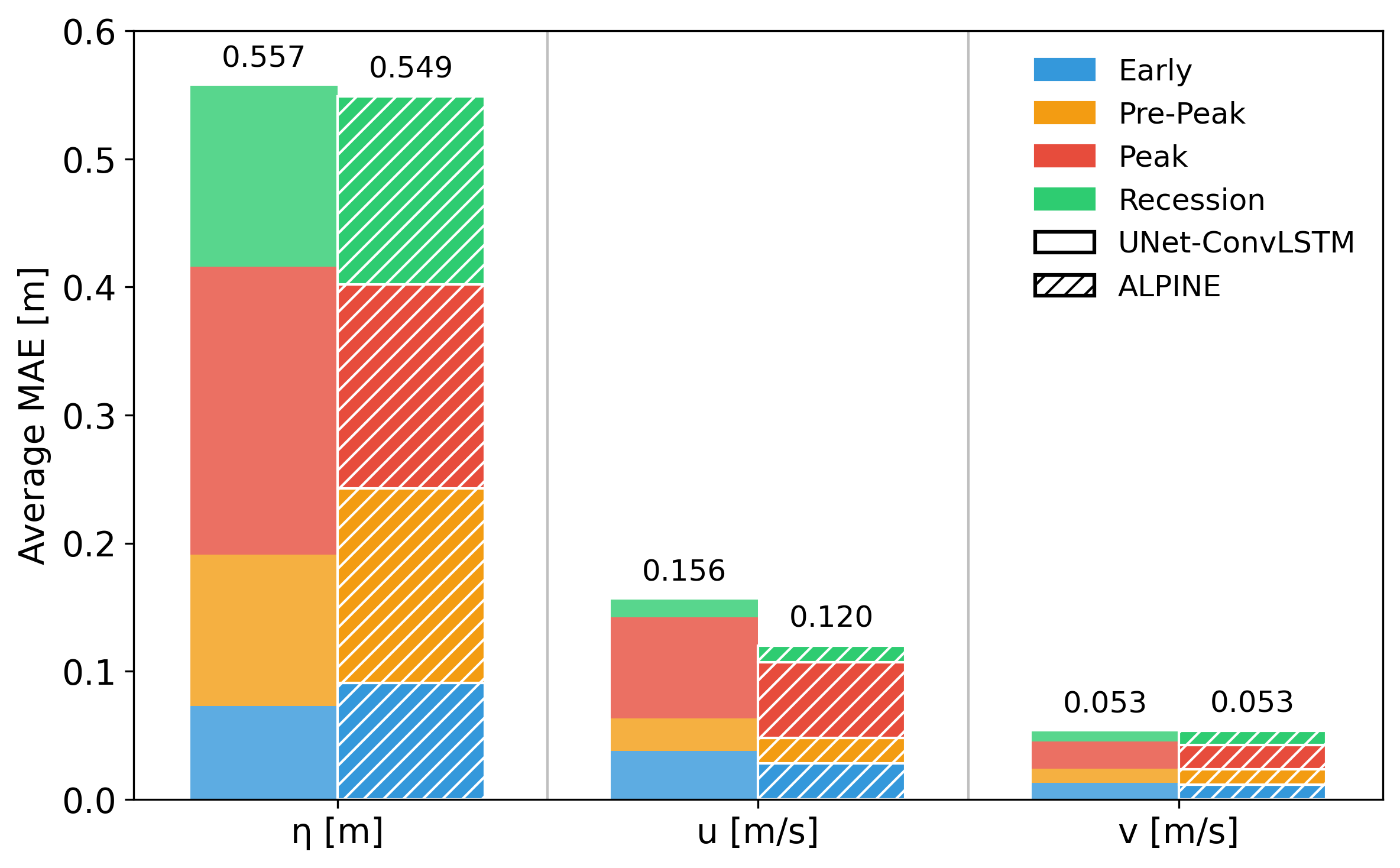}
    \caption{Stacked bar chart comparing mean absolute error (MAE) between UNet-ConvLSTM (solid) and ALPINE (hatched) models across different storm phases during Hurricane Francine. Each bar represents the total MAE for water surface elevation ($\eta$), eastward velocity ($u$), and northward velocity ($v$) predictions, with colors indicating storm phases: Early (blue), Pre-Peak (orange), Peak (red), and Recession (green). The numerical values above each bar show the total domain-averaged MAE across all time steps within each variable category. Storm phases are defined based on Hurricane Francine's temporal evolution: Early (September 4-7), Pre-Peak (September 9), Peak (September 12), and Recession (September 14).
    }
    \label{fig:mae_phase}
\end{figure}

Figure~\ref{fig:mae_phase} presents the domain-averaged mean absolute error (MAE) aggregated across storm phases, providing a complementary perspective to the continuous temporal analysis. MAE is calculated as follows:

\begin{equation}
\mathrm{MAE} = \frac{1}{N} \sum_{i=1}^{N} \left| O_i - P_i \right|
\label{eq:mae}
\end{equation}

MAE measures the average magnitude of prediction errors, and like RMSE, lower values indicate better performance. The stacked bar visualization reveals how prediction errors accumulate during different hydrodynamic regimes during Hurricane Francine.

For $\eta$ predictions, the total MAE across all phases reaches 0.557 m for UNet-ConvLSTM compared to 0.549 m for ALPINE, representing only a modest overall improvement. However, this aggregate metric masks important phase-specific differences. ALPINE demonstrates its greatest advantage during peak storm conditions (red segments), where the physics-informed constraints prove most valuable for maintaining water level accuracy when multiple flood drivers interact. The early and pre-peak phases show comparable performance between models, while the recession phase contributes relatively little to total error for both approaches.

The $u$ velocity component shows more pronounced differences, with ALPINE achieving substantially lower total MAE (0.120 m/s versus 0.156 m/s). The improvement spans all storm phases but is most notable during the early and peak periods when wind-driven currents intensify. The consistent reduction across phases reflects ALPINE's superior momentum conservation, which becomes increasingly important as flow velocities amplify throughout the storm progression. For $v$ predictions, both models achieve similar total MAE values (0.053 m/s), with the phase-specific contributions showing comparable magnitudes across all storm stages. Detailed temporal breakdowns of these metrics are provided in Table~\ref{tab:mae_storm_phases}. The spatial evolution of prediction accuracy for $\eta$, $u$, and $v$ at five representative time steps is presented in Figs.~S7--S9 in the supplementary material.

\begin{table}[ht]
\caption{Domain-averaged mean absolute error (MAE) comparison across storm phases for ALPINE and UNet-ConvLSTM models.}
\centering
\begin{tabular}{clccc}
\toprule
\textbf{Variable} & \textbf{Timestep} & \textbf{UNet-ConvLSTM} & \textbf{ALPINE} & \textbf{Improvement (\%)} \\
\midrule
\multirow{5}{*}{$\eta$} 
& 04 Sep 2024 16:00 & 0.062 & 0.064 & -3.17 \\
& 07 Sep 2024 06:00 & 0.084 & 0.118 & -33.66 \\
& 09 Sep 2024 20:00 & 0.118 & 0.152 & -25.19 \\
& 12 Sep 2024 10:00 & 0.225 & 0.159 & 34.38 \\
& 14 Sep 2024 23:00 & 0.141 & 0.147 & -4.17 \\
\midrule
\multirow{5}{*}{$u$} 
& 04 Sep 2024 16:00 & 0.058 & 0.044 & 27.45 \\
& 07 Sep 2024 06:00 & 0.018 & 0.012 & 40.00 \\
& 09 Sep 2024 20:00 & 0.025 & 0.020 & 22.22 \\
& 12 Sep 2024 10:00 & 0.079 & 0.059 & 28.99 \\
& 14 Sep 2024 23:00 & 0.014 & 0.013 & 7.41 \\
\midrule
\multirow{5}{*}{$v$} 
& 04 Sep 2024 16:00 & 0.016 & 0.015 & 6.45 \\
& 07 Sep 2024 06:00 & 0.010 & 0.008 & 22.22 \\
& 09 Sep 2024 20:00 & 0.011 & 0.012 & -8.69 \\
& 12 Sep 2024 10:00 & 0.021 & 0.019 & 10.00 \\
& 14 Sep 2024 23:00 & 0.008 & 0.011 & -31.58 \\
\bottomrule
\end{tabular}
\label{tab:mae_storm_phases}
\end{table}

Overall, the temporal analysis reveals that ALPINE's physics-informed constraints provide the most significant advantages during periods of rapid hydrodynamic change, particularly when multiple flood drivers interact simultaneously. The volume conservation constraints appear especially beneficial during peak storm conditions when large water volumes are being redistributed across the domain through complex momentum exchanges between riverine inflows, wind-driven setup, and tidal processes. This suggests that the physics-informed approach is most valuable precisely when traditional data-driven models struggle most, i.e., during extreme events that push the system beyond the range of typical training conditions. The error evolution patterns also highlight the challenge both models face in accurately predicting the precise timing and spatial extent of wetting and drying processes. However, ALPINE's enforcement of mass and momentum conservation helps maintain more physically consistent predictions even when local errors occur, preventing the accumulation of non-physical artifacts that can build up over time in unconstrained networks.

\section{Conclusions}
This study presents the first PINN framework to enforce complete shallow water dynamics for CF modeling, integrating mass conservation and two momentum equations within a unified deep learning architecture. The developed PINN model (a.k.a. ALPINE) addresses critical limitations of existing approaches by combining the physical consistency of traditional hydrodynamic models with the computational efficiency and pattern recognition capabilities of modern neural networks. Through comprehensive training and validation on five historical hurricanes and blind testing on Hurricane Francine (2024) in Mobile Bay, Alabama, the proposed framework demonstrates superior generalizability compared to purely data-driven approaches, outperforming the UNet-ConvLSTM baseline particularly during extreme flooding conditions when multiple drivers interact simultaneously.

The physics-informed constraints prove most valuable precisely when reliable predictions are needed most. During Hurricane Francine’s peak intensity (12 September 2024 10:00 in Figure~\ref{fig:temporal_metrics} and Figure~\ref{fig:mae_phase}), ALPINE maintained realistic flood patterns with domain-average RMSE reductions (9\% for $\eta$, 25\% for $u$, and 5\% for $v$ in Table~\ref{tab:summary_performance}) compared to purely data-driven models. This model successfully captures the intricate interactions between storm surge, riverine discharge, and precipitation that characterize CF events, maintaining mass and momentum balance throughout the prediction horizon. Spatial analysis reveals that physics-informed constraints provide the greatest benefits in hydraulically complex regions where multiple flood drivers converge, such as river-bay junctions and tidal channel networks. Most importantly, the enforcement of conservation laws prevents the accumulation of non-physical artifacts that commonly arise in traditional neural networks during autoregressive prediction. This physics-based foundation enables the framework to maintain realistic predictions even when extrapolating to extreme conditions and unseen storm scenarios, demonstrating enhanced generalizability compared to purely data-driven surrogates that typically fail beyond their training conditions.

The trained PINN achieves inference through a single forward pass through the network, enabling it to harness modern GPUs for near-real-time forecasts without sacrificing physical consistency. This capability, together with its demonstrated skill during an unseen event featuring concurrent river discharge and wind-driven surge, positions physics-informed surrogates as a practical tool for operational compound-flood warning systems, emergency management decision support systems, and large-scale ensemble runs required for probabilistic risk assessment. Future developments expand this framework through distributed computing frameworks, multi-GPU parallelism, and model compression technique to enhance computational performance and support sub-meter resolution and three-dimensional flow representation \citep{Huang2025, Shukla2021}. The framework could also be broadened through (i) multi-basin transfer learning \citep{Daramola2025, Xu2023} to enable a single network to handle diverse coastal morphologies, (ii) probabilistic PINN formulations to quantify predictive uncertainty \citep{Psaros2023, Shih2025}, and (iii) direct coupling with atmospheric and hydrologic forcings for fully end-to-end forecasting

\paragraph{CRediT authorship contribution statement}
\textbf{Soheil Radfar:} Conceptualization, Data curation, Formal analysis, Investigation, Methodology, Software, Validation, Visualization, Writing – original draft. \textbf{Faezeh Maghsoodifar:} Data curation, Methodology, Software, Visualization, Writing – original draft. \textbf{Hamed Moftakhari:} Conceptualization, Funding acquisition, Resources, Supervision, Validation, Writing – review and editing. \textbf{Hamid Moradkhani:} Funding acquisition, Resources, Supervision, Writing – review and editing.
\paragraph{Declaration of competing interest}
The authors declare that they have no known competing financial interests or personal relationships that could have appeared to influence the work reported in this paper.
\paragraph{Acknowledgement}
This research was supported by the Cooperative Institute for Research to Operations in Hydrology (CIROH) with funding under award NA22NWS4320003 from the NOAA Cooperative Institute Program. The statements, findings, conclusions, and recommendations are those of the author(s) and do not necessarily reflect the opinions of NOAA.

\end{document}